\newlength{\saveparindent}
\newlength{\saveparskip}
\documentclass[letterpaper,11pt]{article}

\usepackage{amsmath}

\usepackage{xspace}
\usepackage{latexsym}
\usepackage{cite}
\usepackage{epsfig}
\usepackage{color}
\usepackage{times}
\usepackage{fullpage}
\renewcommand{\epsilon}       {\varepsilon}

\newtheorem{theorem}        {Theorem}
\newtheorem{lemma}[theorem] {Lemma}
\newtheorem{corollary}[theorem] {Corollary}
\newtheorem{xample}     {Example}

\def\squarebox#1{\hbox to #1{\hfill\vbox to #1{\vfill}}}

\newcommand{\qedbox}        {\vbox{\hrule\hbox{\vrule\squarebox{.667em}\vrule}\hrule}}
\newcommand{\qed}       {\nopagebreak\mbox{}\hfill\qedbox\smallskip}

\let\latexcite=\cite
\def\cite{\nolinebreak\latexcite}
\let\latexref=\ref
\def\ref{\nolinebreak\latexref}

\newcommand{\secref}[1]     {Section~\ref{sec:#1}}

\newcounter{ccount}

\makeatletter
\renewcommand\section{\@startsection {section}{1}{\z@}%
                                   {-3.5ex \@plus -1ex \@minus -.7ex}%
                                   {2.3ex \@plus.2ex}%
                                   {\normalfont\Large\bfseries}}
\makeatother

\begin{document}

\newcommand{\IS}{\textsc{Insertion Sort}\xspace}
\newcommand {\GIS}{\textsc{Gapped Insertion Sort}\xspace}
\newcommand {\LS}{\textsc{Library Sort}\xspace}
\newcommand {\LSC}{\textsc{Library Sort:}\xspace}

\newcommand{\mytitle}{}

\title{\textbf{\IS is \boldmath$O(n\log n)$}
   \thanks{To appear in Proceedings of the Third International Conference on
Fun With Algorithms, FUN 2004.}}

\author{
    Michael A. Bender%
    \thanks{Department of Computer Science, SUNY Stony Brook, Stony Brook, NY 11794-4400, USA;
        \texttt{bender@cs.sunysb.edu}.}
    \and
    Mart\'{\i}n~Farach-Colton%
    \thanks{Department of Computer Science, Rutgers University, Piscataway, NJ 08854, USA;
    \texttt{farach@cs.rutgers.edu}.}%
    \and
    Miguel Mosteiro%
    \thanks{Department of Computer Science, Rutgers University, Piscataway, NJ 08854, USA;
    \texttt{mosteiro@cs.rutgers.edu}.}%
       }

\maketitle


\begin{abstract}

  Traditional \IS runs in $O(n^2)$ time because each insertion takes
  $O(n)$ time.  When people run \IS in the physical world, they leave
  gaps between items to accelerate insertions.  Gaps help in computers
  as well.  This paper shows that \GIS has insertion times of $O(\log
  n)$ with high probability, yielding a total running time of $O(n\log
  n)$ with high probability.

\end{abstract}


\section*{Keywords}
Sorting, Library Sort, Insertion Sort, Gapped Insertion
Sort. ACM-class:~F.2.2,~E.5. arXiv:~cs.DS/0407003. CoRR~Subj-class:~DS-Data Structures and Algorithms.

\section{Introduction}
\label{sec:intro}

Success has its problems.  While many technology companies are
hemorrhaging money and employees, Google is flush with money and
hiring vigorously.  Google employees are cheerful and optimistic,
with the exception of G---.

G--- maintains the mailboxes at Google.  The mailbox technology
consist of trays arranged in linear order and
bolted to the wall.  The names on the mailboxes are alphabetized.
G--- is grumpy after each new hire because, to make room for the
$n$th new employee, the names on $O(n)$ mailboxes need to be
shifted by one.

University graduate programs in the US have also been growing
vigorously, accepting as students those talented employees downsized
from high-tech companies.

At Stony Brook S--- implements the mailbox protocol.  Each time a
new student arrives, S--- makes room for the new
student's mailbox using the same technique as G---. However, S---
only needs to shift names by one until a gap is reached, where the
empty mailbox belonged to a student who graduated previously.
Because the names have  more or less random rank, S--- does not
need to shift many names before reaching a gap.

Both S--- and G--- are implementing \IS.  However, while S---
is blissfully unaware that \IS is an $O(n^2)$ algorithm,
G--- continually hopes that each new hire will be named Zhang, Zizmor,
or Zyxt.

R---, the librarian at Rutgers, is astonished by all the fuss over
insertions.  R--- inserts new books into the
stacks\footnote{Throughout, we mean \emph{library stacks}, an ordered
  set of stacks, rather than lifo stacks.} every day.  R--- plans for
the future by leaving gaps on every shelf.  Periodically, R--- adds
stacks to accommodate the growing collection.  Although spreading the
books onto the new stacks is laborious, these rearrangements happens
so infrequently that R--- has plenty of time to send overdue notices
to hard-working students and professors.

This paper shows that \GIS, or \LS, has insertion times of $O(\log
n)$ with high probability, yielding a total running time of $O(n\log
n)$ with high probability.

\subsection*{Standard \IS}
In standard \IS we maintain an array of elements in sorted order.
When we insert a new element, we find its target location and
slide each element after this location ahead by one array position to
make room for the new insertion.  The $i$th insertion takes time
$O(i)$, for a total of $O(n^2)$.  Finding the target position of the
$i$th element takes time $O(\log i)$ using binary search, though this
cost is dominated by the insertion cost.

\subsection*{\LS}
We achieve $O(\log n)$-time insertions with high probability by
keeping gaps evenly distributed between the inserted elements and
randomly permuting the input.  Then we only need to move a small
number of elements ahead by one position until we reach a gap.  The
more gaps we leave, the fewer elements we move on insertions.
However, we can tolerate a small-constant-space overhead.

The remainder of this paper is organized as follows. We present the
details of the algorithm in \secref{algorithm} and show in
\secref{analysis} that the algorithm runs in $O(n\log n)$ time with
high probability.  In \secref{conclude} we conclude with a few
comments and some related work.


\section{\LSC Algorithm and Terminology}
\label{sec:algorithm}

Let $A$ be an $n$-element array to be sorted.  These elements are
inserted one at a time in random order into a \emph{sorting array}
$S$ of size $(1+\epsilon)n$.  The insertions proceed in $\log n$
\emph{rounds} as follows.  Each round doubles the number of elements
inserted into $S$ and doubles the prefix of $S$ where elements reside.
Specifically, round $i$ ends when element $2^{i}$ is inserted and the
elements are \emph{rebalanced}.  Before the rebalance, the $2^i$
elements are in the first $(1+\epsilon)2^i$ positions.  A rebalance
moves them into the first $(2+2\epsilon)2^i$ positions, spreading the
elements as evenly as possible.  We call $2+2\epsilon$ the
\emph{spreading factor}.

During the $i$th round, the $2^{i-1}$ elements in $S$ at the beginning of
the round are called \emph{support elements}, and their initial
positions are called \emph{support positions}.  The $2^{i-1}$
elements inserted before the end-of-round rebalance are called
\emph{intercalated elements}.

The insertion of $2^{i-1}$ intercalated elements within round $i$ is
performed the brute force way: search for the target position of the
element to be inserted by binary search (amongst the $2^{i-1}$ support
positions in $S$), and move elements of higher rank to make room for
the new element.  Not all elements of higher rank need to be moved,
only those in adjacent array positions until the nearest gap is found.


\section{Analysis}

\label{sec:analysis}

For the sake of clarity, we divide the time complexity into four
parts: the rebalance cost, the search cost, the insertion cost for the
first $\sqrt n$ elements, and the insertion cost for the remaining
elements.  Let $m$ be the number of elements inserted at any time.

\subsection{Insertion cost for \boldmath$m=O(\sqrt n)$, rebalance cost,
and search cost}

\begin{lemma}
\label{firstins}
The insertion time for the first $O(\sqrt n)$ insertions is $O(n)$.
\end{lemma}
\begin{proof}
By the quadratic running time of \IS.\qed
\end{proof}

\begin{lemma}
\label{rebal}
For a given input of size $n$, the cost of all rebalances is $O(n)$.
\end{lemma}
\begin{proof}
  Since the number of elements doubles in each round and the spreading
  factor is constant, the cost of spreading $m$ elements on each
  rebalance is amortized over the previous $m/2$ insertions, for an
  amortized rebalance cost of $O(1)$ per insertion.\qed
\end{proof}

\begin{lemma}
\label{search}
The cost of finding the location to insert a new element in the
sorting array is $O(\log m)$.
\end{lemma}
\begin{proof}
  Binary search among the $O(m)$ support positions takes time
  $O(\log m)$.  A final search between two support positions takes
  time $O(1)$, since the spreading factor is constant.\qed
\end{proof}

\subsection{Insertion cost for  \boldmath$m=\Omega(\sqrt n)$}

We now bound the number of elements moved per insertion when
$m=\Omega(\sqrt n)$ elements
have already been inserted. We show that with high
probability, for sufficiently large $c$, all sets of $c\log m$
contiguous support elements have fewer than $(1+\epsilon)c\log m$
intercalated elements inserted among them by the end of the round.
The $c\log m$ support elements are spread among $(2+2\epsilon)c\log m$
sorting array positions at the beginning of a round.  Therefore, after
$(1+\epsilon)c\log m$ intercalated elements are added, there will
still be gaps --- indeed, there will be $\epsilon c\log m$ empty array
positions.  Thus, each insertion takes time $O(\log m)$ because
shifts propagate until the next gap, which appears within $O(\log
m)$ positions.  This observation establishes our result.

\subsection*{The direct approach}
Let $D$ be a set of $c\log m$ contiguous support elements.  We would
like to compute the number of intercalated elements that land among
the elements of $D$.  Notice that if there are $k$ elements in the
sorting array, then the $k+1$st intercalated element is equally likely
to land between any of those $k$ elements.  Thus, if an intercalated
element is inserted within $D$, the probability of further
insertions within $D$ increases, and conversely, if an intercalated
element is inserted outside of $D$, the probability of
further insertions within $D$ decreases.

We formalize the problem as follows.  Consider two urns, urn $A$
starting with $c\log m$ balls and urn $B$ starting with $m-c\log m$
balls.  Throw $m$ additional balls, one after another, into one of the
two urns with probability proportional to the number of balls in each
urn.  Let random variable $X_i = 1$ if ball $i$ lands in urn $A$ and
let $X_i=0$ if ball $i$ lands
in urn $B$.  We now need to bound the tails of $\sum X_i$.

Because these $X_i$ are positively correlated, bounding the tail of
their sum is awkward.  We analyze a simpler game below.

\subsection*{The arrival permutation}
We first set up the problem.  Consider $2m$ elements to sort.  Each of
the $(2m)!$ orders of insertion is equally likely.  We refer to each
insertion order as an \emph{arrival permutation}.  The first half of
the arrival permutation consists of support elements, and the second
half consists of intercalated elements.  Thus, the probability of
being a support (resp.~intercalated) element equals the probability of
being in the first (resp.~second) half of the arrival permutation.

Our goal is to show that for sufficiently large $c$, with high
probability in every set of $(2+\epsilon)c\log m$ contiguous elements,
there are at least $c\log m$ support elements and at least $c\log m$
intercalated elements at the end of a round.  Thus, with high
probability there are also at most $(1+\epsilon)c\log m$ of each type
in every set of $(2+\epsilon)c\log m$ contiguous elements.  Because
the at least $c\log m$ support elements are spread out in a subarray
of size $(2+2\epsilon)c\log m$, there is room to add the at most
$(1+\epsilon)c\log m$ intercalated elements while still leaving gaps.
Therefore, with high probability no insertion will move more than
$(2+\epsilon)c\log m$ elements.

\begin{theorem}
\label{main}
In any set $C$ of $(2+\epsilon)c\log m$ contiguous elements, there are
at least $c\log m$ support elements and at least $c\log m$
intercalated elements with high probability.
\end{theorem}
\begin{proof}
  \newcommand{\AP}{\mbox{${\cal P}$}\xspace} Consider choosing an
  arrival permutation \AP of length $2m$ uniformly at random by placing
  the elements one-by-one into \AP, selecting an empty slot uniformly
  at each step.  We place the elements of set $C$ into \AP before
  placing the elements of $\overline{C}$ in \AP.
  We give an upper bound on the number of elements in $C$
  that are support elements, that is, the number of elements
  that fall in the first half of \AP.
  The argument for intercalated elements is symmetric.

  Let $s_i$ be the number of elements already inserted into the first half
  of \AP just before the $i$th insertion.  The probability $p_i$ that
  the $i$th insertion is in the first half of \AP is then
  $(m-s_i)/(2m-i+1)$.

  Let random variable
  $X_i=1$ if element $i$ is a support element, and let $X_i=0$ otherwise.
  Random variables $X_1,\ldots,X_{2m}$ now depend on the remaining ]
  blank spaces in the
  permutation and are negatively correlated.  Furthermore,
  $|C|=(2+\epsilon)c\log m$ is small compared to $m$, and so $E[X_i] =
  p_i$ is very close to $1/2$ for the first $|C|$ element.  Given this
  bound, we can prove our theorem with a straightforward application
  of Chernoff bounds.

  Here we prove the theorem using elementary methods, as
  follows. The probability that a given element in
  $C$ is a support element is at most
  $$\frac{m}{2m-|C|+1} = \frac{m}{2m-(2+\epsilon)c\log m+1}$$
  Let $p$ be the probability that there are
  at most $c\log m$ support elements in the set $C$.  Then,

\begin{eqnarray}
p &\leq&
\sum_{j=0}^{c\log m}
\binom{|C|}{j}
\left(\frac{m}{2m-|C|+1}\right)^{j}
\left(\frac{m}{2m-|C|+1}\right)^{|C|-j}\nonumber \\
 &\leq&
\left(\frac{m}{2m-|C|+1}\right)^{|C|}\,\,
\sum_{j=0}^{c\log m}
\binom{|C|}{j}. \nonumber
\end{eqnarray}

Bounding the summation by the largest term, we obtain

\[
p \leq
\left(\frac{m}{2m-|C|+1}\right)^{|C|}
c\log m
{\binom{(2+\epsilon)c\log m}{c\log m}}.
\]

Using Stirling's approximation, for some constant $c'$ we obtain

$$
p \leq c'
\left(\frac{m}{2m-|C|+1}\right)^{|C|}
\sqrt{c\log m}
\left(\frac{(2+\epsilon)^{(2+\epsilon)}}{(1+\epsilon)^{(1+\epsilon)}}\right)^{c\log
m}.$$

Manipulating on the first term, we obtain
$$
p \leq c'
\left(1 + \frac{|C|}{m}\right)^{|C|}\left(\frac{1}{2}\right)^{|C|}
\sqrt{c\log m}
\left(\frac{(2+\epsilon)^{(2+\epsilon)}}{(1+\epsilon)^{(1+\epsilon)}}\right)^{c\log
m}.$$

Since $|C|\ll m$, we replace the first term by a constant less than $e$
(indeed, $1+o(1)$) and fold it, along with $c'$, into $c''$.  We get

$$
p \leq c''
\left(\frac{1}{2}\right)^{|C|}
\sqrt{c\log m}
\left(\frac{(2+\epsilon)^{(2+\epsilon)}}{(1+\epsilon)^{(1+\epsilon)}}\right)^{c\log
m}.$$

Since $|C|=(2+\epsilon)c\log m$, the previous equation simplifies to

$$
p \leq c''
\sqrt{c\log m}
\left(\frac{(2+\epsilon)^{(2+\epsilon)}}{2^{(2+\epsilon)}(1+\epsilon)^{(1+\epsilon)}}\right)^{c\log
m}.$$

This last factor is 1 when $\epsilon = 0$, and decreases with
increasing $\epsilon$.  Thus, for any constant $\epsilon > 0$,
the probability $p$ is polynomially
small when $m$ is $\Omega(\sqrt{n})$.  The same result can be
symmetrically obtained for intercalated elements; thus, we have at
least $c\log m$ elements of each, with high probability.\qed
\end{proof}

Note that since $C$ is split evenly in expectation, the expected
insertion cost is constant.

\begin{corollary}
\label{corol}
There are at least $c\log m$ support elements and at least $c\log m$
intercalated elements in each set of $(2+\epsilon)c\log m$ contiguous
elements with high probability.
\end{corollary}
\begin{proof}
  We are interested only in nonoverlapping sets of contiguous
  elements and there are $m/(2+\epsilon)c\log m$ such sets.
  By Theorem~\ref{main} and the union bound, the claim holds.\qed
\end{proof}

\subsection{Summary}
We summarize our results as follows: The overall cost of rebalancing
is $O(n)$ by Lemma~\ref{rebal}. By Lemma~\ref{search}, the cost of
searching for the position of the $i$th element is $O(\log i)$, and
then the overall searching cost is $O(n \log n)$. We proved in
Lemma~\ref{firstins} that the insertion cost of the first $O(\sqrt n)$
elements is $O(n)$ in the worst case. Finally, Theorem~\ref{main}
shows that for sufficiently large $c$, no contiguous $c\log m$
elements in the support have $(1+\epsilon)c\log m$ intercalated
elements inserted among them at the end of any round, with high
probability.  Thus, there is a gap within any segment of
$(2+\epsilon)c\log m$ elements with high probability.  Therefore, the
insertion cost per element for $m=\Omega(\sqrt n)$ is $O(\log m)$ with
high probability.  The overall cost of \LS is $O(n\log n)$ with high
probability.


\section{Conclusions and related work}
\label{sec:conclude}

We have shown that \LS outperforms traditional \IS. There is a
trade-off between the extra space used and the insertion time
given by the relation between $c$ and $\epsilon$. The lower the
desired insertion cost, the bigger the required gap between
elements at rebalance.

\LS is based on the priority queue presented in Itai, Konheim, and
Rodeh~\cite{IKR}. Our analysis is a simplification of theirs.
Moreover, we give high probability and expectation bounds for the
insertion cost, whereas they only give expectation bounds.

An algorithm similar to \LS was presented by Melville and Gries
in~\cite{MG}. This algorithm has a $1/3$ space overhead as compared
with the $\epsilon$ space overhead of \LS. They point out that their
running time analysis was too complicated to be included in the
journal version. To quote the authors: \emph{``We hope others may
develop more satisfactory proofs.''}.

The idea of leaving gaps for insertions in a data structure is
used by Itai, Konheim, and Rodeh~\cite{IKR}.  This idea has found
recent application in external memory and cache-oblivious algorithms in
the \emph{packed memory structure} of Bender, Demaine and
Farach-Colton~\cite{BDF} and later used in~\cite{BDIW,BFJ,F}.


\section*{Acknowledgments}
We would like to thank Erik Demaine and Ian Munro for helpful
discussions. MAB was supported in part by the Sandia National
Laboratories and NSF Grants ACI-032497, CCR-0208670, and EIA-0112849.
MFC was supported in part by CCR-9820879.

\end{document}